\documentclass[fleqn,usenatbib]{mnras}

\usepackage{newtxtext,newtxmath}

\usepackage[T1]{fontenc}
\usepackage{ae,aecompl}

\usepackage{graphicx}	
\usepackage{amsmath}	
\usepackage{amssymb}	
\usepackage{cprotect}

\usepackage{etoolbox}
\makeatletter
\patchcmd\@combinedblfloats{\box\@outputbox}{\unvbox\@outputbox}{}{%
  \errmessage{\noexpand\@combinedblfloats could not be patched}%
}%
\makeatother

\title[The 2018 outburst of BHXB H1743$-$322 as seen with MeerKAT]{The 2018 outburst of BHXB H1743$-$322 as seen with MeerKAT}

\author[D.R.A. Williams et al.]{D.R.A. Williams,$^{1}$\thanks{E-mail: david.williams@physics.ox.ac.uk}
S.E. Motta$^{1}$,
R. Fender$^{1,2}$,
J. Bright$^{1}$,
I. Heywood$^{1,3}$, 
E. Tremou$^{4}$,
\newauthor{P. Woudt$^{2}$,
D.A.H. Buckley$^{5}$,
S. Corbel$^{4,6}$,
M. Coriat$^{7}$,
T. Joseph$^{8}$,
L. Rhodes$^{1,9}$,
}
\newauthor{G.R. Sivakoff$^{10}$,
A.J. van der Horst$^{11,12}$
}
\\
$^{1}$Department of Physics, University of Oxford, Denys Wilkinson Building, Keble Road, Oxford, OX1 3RH, UK \\
$^{2}$Department of Astronomy, University of Cape Town, Private Bag X3, Rondebosch 7701, South Africa \\
$^{3}$Department of Physics and Electronics, Rhodes University, PO Box 94, Grahamstown 6140, South Africa\\
$^{4}$Laboratoire AIM (CEA/IRFU - CNRS/INSU - Universit\'{e} Paris Diderot), CEA DSM/IRFU/SAp, F-91191 Gif-sur-Yvette, France\\
$^{5}$South African Astronomical Observatory, PO Box 9, Observatory 7935, Cape Town, South Africa\\
$^{6}$Station de Radioastronomie de Nan\c{c}ay, Observatoire de Paris, PSL Research University, CNRS, Univ. Orl\'{e}ans, 18330 Nan\c{c}ay, France\\
$^{7}$IRAP, Universite de Toulouse, CNRS, CNES, UPS, 9 av. du colonel Roche, F-31038, Toulouse, France \\
$^{8}$Department of Physics and Astronomy, University of Manchester, Oxford Road, Manchester, M13 9PL, UK \\
$^{9}$Max-Planck-Institut f{\"u}r Radioastronomie, Auf dem H{\"u}gel 69, D-53121 Bonn, Germany\\
$^{10}$Department of Physics, University of Alberta, CCIS 4-181, Edmonton, AB T6G 2E1, Canada\\
$^{11}$Department of Physics, the George Washington University, 725 21st Street NW, Washington, DC 20052, USA\\
$^{12}$Astronomy, Physics and Statistics Institute of Sciences (APSIS), 725 21st Street NW, Washington, DC 20052, USA\\
}

\vspace{-0.5cm}
\date{Accepted XXX. Received YYY; in original form ZZZ}

\pubyear{2019}

\begin{document}
\label{firstpage}
\pagerange{\pageref{firstpage}--\pageref{lastpage}}
\maketitle

\vspace{-0.5cm}
\begin{abstract}
In recent years, the black hole candidate X-ray binary system H1743$-$322 has undergone outbursts and it has been observed with X-ray and radio telescopes. We present 1.3\,GHz MeerKAT radio data from the ThunderKAT Large Survey Project on radio transients for the 2018 outburst of H1743$-$322. We obtain seven detections from a weekly monitoring programme and use publicly available \textit{Swift} X-ray Telescope and \textit{MAXI} data to investigate the radio/X-ray correlation of H1743$-$322 for this outburst. We compare the 2018 outburst with those reported in the literature for this system and find that the X-ray outburst reported is similar to previously reported `hard-only' outbursts. As in previous outbursts, H1743$-$322 follows the `radio-quiet' correlation in the radio/X-ray plane for black hole X-ray binaries, and the radio spectral index throughout the outburst is consistent with the `radio-quiet' population.
\end{abstract}



\vspace{-0.5cm}
\section{Introduction}

Black hole (BH) X-ray binaries (BHXBs) are systems that consist of a BH orbited by a 
stellar companion, typically a low-mass, evolved star, which transfers mass 
by Roche Lobe overflow via an accretion disk. 
BHXBs are typically discovered in outburst, when the X-ray flux increases by several orders of magnitude \citep{RemillardMcClintock2006}, which is also seen at other wavelengths. BHXB outbursts exhibit a number of different states, defined by their X-ray spectral and timing properties \citep[e.g.][for relevant reviews]{RemillardMcClintock2006, BelloniMotta}, and are associated with varying levels of radio emission \citep{Corbel2000,FenderBelloni,FenderBelloniGallo}. 

At the start of an outburst, BHXBs reside in the X-ray `hard' state at low X-ray fluxes, dominated by non thermal emission thought to arise from Compton up-scattering of seed photons from the accretion disk. The hard state is associated with a flat radio spectrum: S $\propto$ $\nu ^{\alpha}$, where S is the flux density in a given frequency band, $\nu$, and $\alpha$ is the spectral index. The radio spectral index is $\sim$0 for a flat spectrum compact radio jet \citep[e.g.][]{Corbel2000,Fender2001,Stirling2001}, analogous to those found in active galactic nuclei \citep[AGN, e.g.][]{Blandford1979}. When a BHXB goes into outburst, the X-ray flux increases by several orders of magnitude over a period of days to weeks \citep{RemillardMcClintock2006}. Usually, after spending between days to months in the hard state, a BHXB may transition to the `soft' state, which is characterised by thermal X-ray emission from an optically-thick, geometrically-thin accretion disk. Radio emission in the soft state is observed to be very low or absent altogether, indicating that the core-jet has been quenched \citep{Fender1999,Corbel2001,Russell2011,Coriat2011}. In between these two states lie the `intermediate states' \textbf{\citep{BelloniMotta}}, where the X-ray spectra evolve from the `hard' to the `soft' states and transient radio ejections of material in the form of relativistic radio jets are observed. These radio ejecta are optically thin ($\alpha <$ 0) and in some sources are observed to travel into the surrounding interstellar medium (ISM) for weeks to years \citep{MirabelRodriguez94,Hjellming95,Hjellming98,corbel2002,Corbel2005,Hannikainen09}.

A connection between the radio and X-ray fluxes in the hard state was discovered for the BHXB GX 339$-$4 and subsequently other BHXBs \citep{Corbel2000,Corbel2003,GalloFenderPooley,Corbel2013}. This non-linear correlation takes the form L$_{\rm Radio}$ $\propto$ L$_{\rm X-ray}^{\sim0.7}$, where L is the luminosity, and suggests a coupling between the accretion flow and the compact jets observed in the hard state. When a third variable, the BH mass, is added to this relationship, this correlation is observed to extend to AGN masses, and forms what is referred to as the "Fundamental Plane of Black Hole Activity" \citep{Merloni2003,Falcke2004,Plotkin2012}, suggesting that the physical processes governing accretion and the generation of jets scale with black hole mass. 

Subsequent observations of more BHXBs indicated a second population of sources with lower radio luminosities, which populate a so-called `radio-quiet' track on the radio/X-ray plane \citep[and references there-in]{GalloFenderPooley,Corbel2004,Coriat2011}. The relationship found in the earlier literature for sources like GX 339$-$4 is now referred to as the `radio-loud' track. The physical mechanism that leads to these two tracks is unknown, although radiative efficiency in the accretion flow \citep{Coriat2011}, inclination effects \citep{Motta2018}, jet magnetic field strength \citep{Casella2009} and the existence of a weak inner disc \citep{Meyer2014} have all been suggested as possible reasons for the two tracks. Although it should be noted that in a statistical sense, the radio/X-ray correlation is consistent with a single population when clustering analysis is used \citep{Gallo2014,Gallo2018}. Thus, the observed tracks in the radio/X-ray plane are important for understanding the differences in the physical coupling between accretion disk and the jet physics in BHXBs.

H1743$-$322 is a black hole candidate that was discovered in 1977 \citep{Doxsey1977,Kaluzienski_Holt1977}, and was one of the first sources discovered to lie on the radio-quiet track \citep{Coriat2011}. It has undergone many complete outbursts since 2003 \citep[e.g. see][]{Coriat2011}, and is also observed to undergo incomplete outbursts, during which only the hard and intermediate \citep{Prat2009, Capitanio2009, Coriat2011}, or hard only \citep{Watchdog} states are sampled.
On MJD 58363.6, H1743$-$322 was reported to be in the early stages of an outburst during \textit{INTEGRAL} observations of the Galactic Center region \citep{2018ATel}; the source was detected at a 7$\sigma$ level in the 20--60 keV band, increased in brightness by the following day and was found to be in the hard state, indicative of a new outburst \citep{2018ATel}. Hence, as part of the ThunderKAT\footnote{\textbf{T}he \textbf{hun}t for \textbf{d}ynamic and \textbf{e}xplosive \textbf{r}adio transients with Meer\textbf{KAT}} Large Survey Project \citep{ThunderKAT2017} with the MeerKAT telescope, we started near-weekly monitoring of this source from MJD 58366, \citep{2018ATel2}, just three days after the first detection of the new outburst, until a month after the end of the outburst on MJD 58432.

\vspace{-0.5cm}

\section{Observations and Data Reduction}
\subsection{MeerKAT radio data of the 2018 outburst}
\label{sec:radioobs}
ThunderKAT observations of H1743$-$322 were performed at a central frequency of 1.284\,GHz, with a bandwidth of 856\,MHz. With the exception of the first observation, each observation consisted of a 15 minute scan of H1743$-$322, interleaved by 2 minute scans of the phase calibrator: either J1712$-$281 (NVSS J171257-280935) or J1830$-$3602 (NVSS J183058-360231), depending on the observation (see Table~\ref{tab:radiodata}). For the initial observation (MJD 58366.734), the target field was observed in four 10 minute scans, interleaved with two minute scans of J1712$-$281. The primary calibrator to set the flux scale and band pass, PKS J1939$-$6342, was observed at the start of each observation for 10 mins.

\begin{table}
	\centering
	\caption{Paired MeerKAT radio (top row) and \textit{Swift}/\textit{MAXI} X-ray (bottom row) observations of H1743$-$322. The radio and X-ray observations were grouped within one day of each other (see Section~\ref{sec:simultaneous}). The table columns are as follows: (1) Central MJD of the radio or X-ray observation, given the observing time (see Section~\ref{sec:radioobs}, the superscript $\alpha$ or $\beta$ denotes the phase calibrator used: $^{\alpha}$ refers to J1712-281 and $^{\beta}$ refers to J1830-3602, whereas the asterisks denote observations that were grouped over a larger period than one day, in order to provide enough photons to extract a spectral fit and as such only the central MJD is shown, see Section~\ref{sec:xrayobs} for further details.; (2) The rms noise level estimated for the radio image, measured in $\mu$Jy/beam; (3) The peak radio flux density at 1.284\,MHz of H1743$-$322 in mJy/beam if detected, otherwise the 3$\sigma$ upper limit to the source is given; (4) The radio spectral index, $\alpha$, estimated from splitting the radio band into two equal bandwidths (see Section~\ref{sec:radioobs}); (5) The 1$-$10\,keV flux given from spectral fitting if detected, if not the upper limit is given, the fluxes are in units of 10$^{-9}$\,erg cm$^{-2}$s$^{-1}$. The superscript `S' or `M' denotes the X-ray instrument used to ascertain the fluxes described in Section~\ref{sec:xrayobs}: `S' denotes \textit{Swift XRT} and `M' denotes \textit{MAXI}. }
	\label{tab:radiodata}
	\begin{tabular}{lcccccc} 
		\hline
		MJD  & rms & Radio & Spectral & Unabsorbed \\
		     & $\mu$Jy/ & Flux   & Index & X-ray Flux \\
		     & beam & mJy/beam   & $\alpha$ & erg cm$^{-2}$s$^{-1}$ \\
		 (1) & (2)      & (3)        & (4)      & (5)\\
		 \hline
		58366.734$^{\alpha}$ & 45  & 0.69$\pm$0.07  & -0.32$\pm$0.03 & -              \\
		58369.728$^{\alpha}$ & 95  & 1.4$\pm$0.1  & -1.0$\pm$0.4 & -              \\
		58368.730*           & -   & -              & - & 1.0$\pm$0.2$^{M}$  \\
		\hline
		58375.705$^{\alpha}$ & 65  & 2.42$\pm$0.06  & -0.4$\pm$0.2 & -              \\
		58376.445            & -   & -              & - & 1.52$\pm$0.08$^{S}$\\		 
		\hline
		58382.685$^{\beta}$  & 34  & 1.69$\pm$0.05  & 0.03$\pm$0.07 & -              \\
		58382.556            & -   & -              & - & 1.31$\pm$0.07$^{S}$ \\
		\hline
		58389.716$^{\beta}$  & 36  & 0.88$\pm$0.05  & -0.1$\pm$0.2 & -              \\
		58388.271            & -   & -              & - & 0.79$\pm$0.04$^{S}$ \\		 
		\hline
		58396.662$^{\beta}$  & 37  & 0.32$\pm$0.08  & -1.0$\pm$0.3 & -              \\
		58397.633            & -   & -              & - & 0.187$\pm$0.009$^{S}$ \\		 
		\hline
		58403.633$^{\beta}$  & 36  & 0.2$\pm$0.1  & -0.8$\pm$0.8 & -              \\
		58403.470            & -   & -              & - & 0.071$\pm$0.004$^{S}$ \\		 
		\hline
		58410.591$^{\beta}$  & 39  & $<$0.11        & - & - \\
		58409.589            & -   & -              & - & 0.021$\pm$0.001$^{S}$ \\
		\hline
		58418.511$^{\beta}$  & 35  & $<$0.11        & - & - \\
		58418.500*           & -   & -              & - & 0.009$^{+0.006}_{-0.005}$ $^{S}$\\
		\hline
		58425.487*$^{\beta}$ & 33  & $<$0.10        & - & - \\
		58425.487            & -   & -              & - & $<$0.46$^{M}$ \\
		\hline
		58432.468*$^{\beta}$ & 34  & $<$0.11        & - & - \\
		58432.468            & -   & -              & - & $<$0.11$^{M}$ \\
		\hline
	\end{tabular}
\end{table}

\begin{figure}
	\includegraphics[width=\columnwidth]{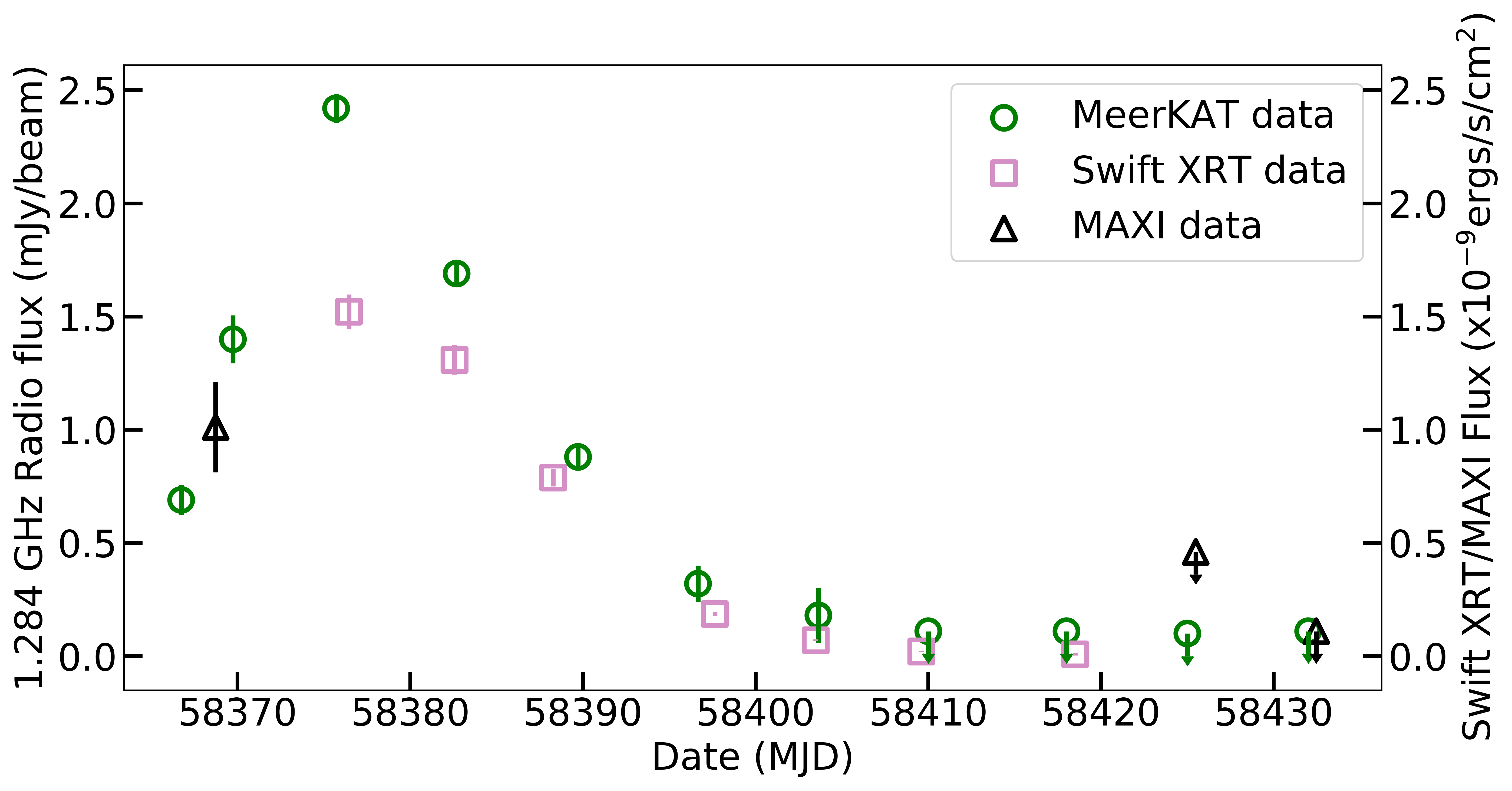}
    \caption{The radio (green circles), \textit{Swift} X-ray (pink squares) and \textit{MAXI} (black triangles) light curves of the 2018 Outburst of H1743$-$322, from the data in Section~\ref{sec:xrayobs} and shown in  Table~\ref{tab:radiodata}. The upper limits are denoted with downward arrows in the MeerKAT and \textit{MAXI} data. The X-ray data are obtained from fitting \textit{MAXI} or \textit{Swift} XRT data in the unabsorbed 1--10\,keV energy range. The radio and X-ray data roughly track with each other, as expected given the radio/X-ray correlation in BHXBs.}
    \label{fig:lc}
\end{figure}

To reduce the MeerKAT radio data, we first flagged any times of radio frequency interference (RFI). We removed the first and last 100 channels (channel width $\sim$209\,kHz) of the band and then clipped visibilities with zero amplitude. We then ran AOFlagger \citep{AOFlagger} to remove further areas of RFI. In general, only around $\sim$20 \textit{per cent} of the data were flagged at this stage. We then calibrated the data using \textsc{CASA} \citep{CASA}. We set the flux scale on the calibrators and then proceeded to solve for phase-only corrections, followed by antenna-based delays. We computed a band pass for PKS J1939$-$6342 and applied these tables to the phase calibrator, solving for the complex gains. Finally, we scaled the gain corrections from the flux calibrator to the phase calibrator and target, and applied all the associated calibration tables to the target field. We performed a small amount of manual flagging at this stage to the target field before imaging with \textsc{wsclean} \citep{wsclean,wsclean2}.

Due to several bright extended sources within the MeerKAT primary beam, we imaged a field 204$\arcmin \times$204$\arcmin$ in size, with a pixel size of 1.5$\arcsec$. The synthesized beam of all the observations was $\sim$5$\arcsec$. We used Briggs weighting with a robust parameter of -0.85 and the auto-thresholding algorithms in \textsc{wsclean} to clean the images. The image rms noise values of each of the datasets are given in Table~\ref{tab:radiodata}. In addition, we produced two images of equal spectral width to investigate the radio spectrum evolution throughout the outburst. However, we note that the band spans only $\sim$400\,MHz between the two data points. 
In order to estimate a flux bootstrapping uncertainty between epochs due to variability in the flux calibrator, we used \textsc{pyBDSF}\footnote{https://www.astron.nl/citt/pybdsf/} software to extract positions and fluxes of all sources in each epoch. We then matched all the unresolved sources and calculated an average standard deviation of 5 \textit{per cent} in the fit values for the sources, which we fold into our flux uncertainties in quadrature in Table~\ref{tab:radiodata}.
\vspace{-0.3cm}
\subsection{X-ray data of the 2018 outburst}
\label{sec:xrayobs}
\subsubsection{\textit{Swift}}\label{sec:swift}
We use the X-ray data obtained from the XRT instrument onboard the Neil Gehrels \textit{Swift} Observatory to extract spectra in the 0.6--10 keV energy band through the \textit{Swift} XRT product generator online reduction pipeline \citep{Evans2007,Evans2009}. We fitted each spectrum in \textsc{XSPEC} with an absorbed power law, with the addition of a Gaussian line centered at 6.4 keV Fe emission when required by the fit. We initially allowed the Galactic neutral hydrogen absorption column parameter, $N_H$, free to vary and fixed the value to the average $N_H$ the fits returned, and then tied the $N_H$ across spectra, assuming that the same $N_H$ value applies to all of them. We find that a $N_H$ slightly different than the Galactic value in the direction of H1743$-$322 (5.5$ \times 10^{22}$\,cm$^{-2}$, see \citealt{HI4PI2016}), though still consistent with previous outbursts is necessary to fit the data, i.e., $N_H$ = 1.99$ \times 10^{22}$\,cm$^{-2}$. We then measured the unabsorbed X-ray fluxes in the 1--10 keV energy band, freezing the value of the photon index, $\Gamma$=1.5, and report these values in Table~\ref{tab:radiodata}. 

\vspace{-0.3cm}
\subsubsection{\textit{MAXI} data}\label{sec:MAXI}

The Monitor of All-sky X-ray Image (\textit{MAXI}) \citep{MAXI} observed H1743$-$322 throughout the 2018 outburst and the daily light curve is publicly available. 
We extracted spectra using the MAXI \textit{on demand} web tool\footnote{http://maxi.riken.jp/mxondem/} by selecting events collected within 3 days from a given MeerKAT observation. This results in a spectrum covering the 2--20 keV energy band, which we fitted in the same way as for the \textit{Swift} spectra using the same $N_H$ value (see Sec. \ref{sec:swift}).

\vspace{-0.3cm}
\subsubsection{Simultaneity of radio and X-ray observations}
\label{sec:simultaneous}
The MeerKAT observations were performed roughly-weekly as part of the ThunderKAT XRB monitoring programme and the \textit{Swift} observations were made approximately every three days. It was not always possible to get simultaneous radio and X-ray coverage with both instruments. As a standard analysis, 
we paired up radio and X-ray observations made within $\pm$1 day of each other \citep[e.g., see ][]{GalloMillerFender}. The only exception is the MeerKAT data on MJD 58389, where the nearest \textit{Swift} data point was obtained 1.5 days later. 
The radio and X-ray light curves are shown in Figure~\ref{fig:lc}.

\vspace{-0.5cm}
\section{Results \& Discussion}
Of the eleven radio observations, H1743$-$322 was detected above the 3$\sigma$ rms noise level in seven. All of the detections show an unresolved point source at the phase centre and no transient radio ejecta were observed. 
The flux density evolved smoothly across the seven detections, from 0.69$\pm$0.07\,mJy/beam on MJD 58366 up to 2.42$\pm$0.06\,mJy/beam on MJD 58375 and was undetected to a 3$\sigma$ noise limit of 110\,$\mu$Jy/beam on MJD 58410. 

The X-ray data from \textit{Swift} XRT show that the X-rays peaked around MJD 58376.44 at 1.52$\pm$0.08$\times$10$^{-9}$\,erg cm$^{-2}$ s$^{-1}$, i.e. within a day of the MeerKAT highest detected flux. 
We note that the source remained in the spectral `hard' state throughout the outburst, as evidenced by the power-law from the spectral fits not exceeding 1.5, giving reason to believe that H1743$-$322 never left the `hard' spectral state in 2018. 
As such, we label the 2018 outburst as `hard-only'. A hard-only outburst explains why no transient radio ejecta were seen in this outburst, contrary to some previous outbursts that underwent a full X-ray spectral evolution \citep[e.g.][]{Corbel2005}. \citet{Watchdog} report that of the thirteen outbursts of H1743$-$322 between the well-studied 2003 outburst and the 2015 outburst, five ($\sim$40 \textit{per cent}) separate outbursts were also found to be `hard-only'. Six of the thirteen outbursts completed a full spectral evolution, and two were found to be indeterminate. Therefore, `hard-only' outbursts are not uncommmon in this source and further investigations need to be undertaken to understand the underlying physical differences between `hard-only' and `successful' outbursts. 

\begin{figure*}
	\includegraphics[width=\textwidth]{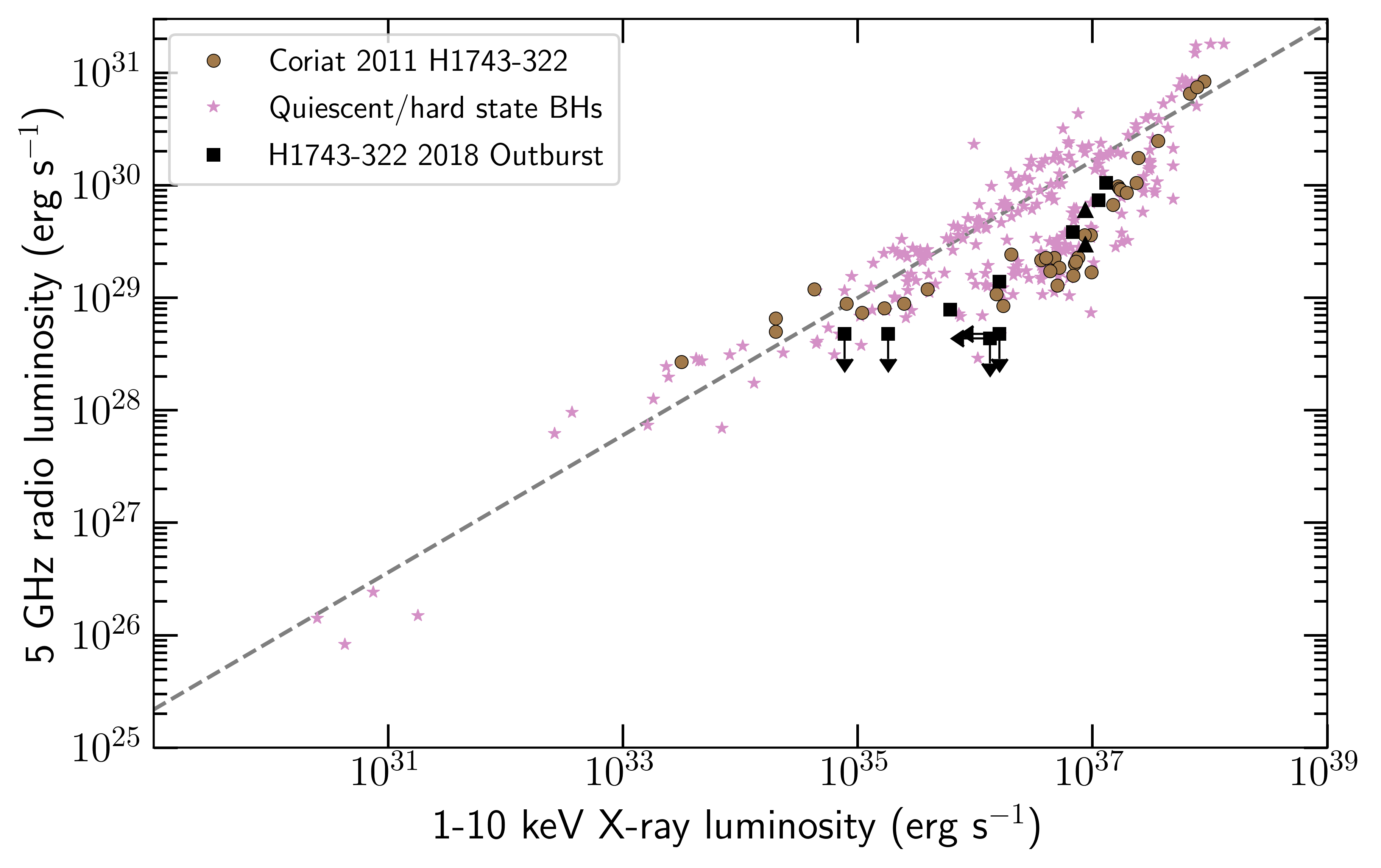}
    \caption{The radio/X-ray plane of the BHXB H1743$-$322 for the 2018 outburst, using the MeerKAT detections with the \textit{Swift} XRT detections (black squares). The data points which are upper limits are denoted with arrows in the relevant axes. The black triangles use an averaged X-ray flux density for the first two MeerKAT detections (see Section~\ref{sec:xrayobs}). The fit for the `radio-loud' track from \citep{arash_bahramian_2018_1252036} is shown as the dashed line. Plotted in brown circles and pink stars are the previous detections of outbursts in H1732$-$322 from the compilation of \citet{Coriat2011} and other BHXBs in \citealt{arash_bahramian_2018_1252036}, respectively. The 2018 outburst of H1743$-$322 is consistent with the compilation data of \citet{Coriat2011}, and the `radio-quiet' track of BHXRBs.}
    \label{fig:LrLx}
\end{figure*}

To compare our MeerKAT observations with the archival radio observations of H1743$-$322 and other hard state BHXBs on the radio/X-ray correlation, we perform a standard analysis and convert the 1.284\,GHz peak flux density for all our observations to 5\,GHz, assuming a flat radio spectral index, to compare directly to the data presented in \cite{arash_bahramian_2018_1252036}. However, we note that our observations indicate steeper radio spectra (e.g. see MJDs 58366/58396), but steeper spectra would result in the source being more radio-quiet than the archival data. We used the 1--10\,keV fluxes from the \textit{Swift} data and \textit{MAXI} data. We assume a distance of 8.5$\pm$0.8\,kpc, estimated from fitting a ballistic jet model to the data of the 2003 outburst \citep{Steiner2012}. Our observations show that for all five contemporaneous radio/X-ray detections, the 2018 outburst follows the `radio-quiet' track (see Fig~\ref{fig:LrLx}). Furthermore, for the first two MeerKAT data points, where only an average X-ray flux over the entire week was available, our first detection was in the middle of the `radio-quiet' track. We do not observe any transition to the radio-loud branch, as previously seen for this source \citep{Coriat2011}.

Recent studies of the radio/X-ray plane have suggested that the sources GX 339$-$4 (the classical `radio-loud' source) and H1743$-$322 (the main `radio-quiet' source) follow different correlations when using a bolometric X-ray flux over a wider X-ray band and when on the rise or decline of an outburst \citep{Islam_Zdziarski2018,Koljonen2019}. Due to the narrow bandpass of the \textit{Swift XRT} and \textit{MAXI} observations shown here and the low number of radio detections, it is not possible to perform these analyses for the 2018 outburst of H1743$-$322, but future outbursts observed with MeerKAT may allow us to investigate these properties further.

Radio-loud and radio-quiet BHXBs are observed to have distinct distributions of radio spectral index ($\alpha$) during outbursts \citep{EspinasseFender} and therefore can be used to investigate the physics of the jet cores in the X-ray spectral hard state. We created two images encompassing half of the MeerKAT pass band in order to compute a two-point in-band radio spectral index of each radio detection during the 2018 outburst. We estimated the spectral index and the associated uncertainty based on these images \citep[see equations in][]{EspinasseFender} and our results are outlined in Table~\ref{tab:radiodata}. Qualitatively, our values of $\alpha$ are similar to those found for other `radio-quiet' sources, i.e. spectrally steep ($\alpha <$ 0). To quantify the likelihood of our values of $\alpha$ being drawn either the radio-quiet or radio-loud distributions, we performed a Kolmogorov-Smirnov (KS) test on our distribution against the Gaussian distributions reported in \citet{EspinasseFender}, with the null hypothesis that the samples are drawn from the same distribution. We note that formally, a KS test requires Gaussian distribution and requires a single peak in the distribution, which our radio spectral indices do not follow. For the radio-loud distribution, we calculated p-value of 2.0$\times$10$^{-4}$, allowing us to reject the null hypothesis. We report a p-value of 0.14 for the radio-quiet distribution, which suggests that we cannot conclude the 2018 outburst of H1743$-$322 is drawn from a different distribution of spectral indices compared to those reported in \citet{EspinasseFender}. From these diagnostics, we conclude that the radio spectra of the 2018 MeerKAT data are consistent with the radio-quiet sources reported in \citet{EspinasseFender}.

\vspace{-0.5cm}
\section{Conclusions}

We followed the 2018 outburst of the black hole candidate X-ray binary H1743$-$322 at 1.284\,GHz with the MeerKAT radio interferometer as part of the long term X-ray binary (XRB) monitoring programme of the Large Survey Project ThunderKAT. In total, we detected H1743$-$322 seven times over the course of the 2018 outburst. We analysed publicly available X-ray data from the \textit{Swift} and \textit{MAXI} X-ray telescopes. We obtained five \textit{Swift} data points with quasi-simultaneous (observations obtained with one day of each other) radio coverage. The remaining two radio data points were supplemented with quasi-simultaneous \textit{MAXI} data. We combined the MeerKAT observations with the X-ray data to investigate the non-linear radio/X-ray correlation for this source against that for other known BHXB systems in the hard state. We found that the 2018 outburst of H1743$-$322 followed the `radio-quiet' track previously shown by this source \citep{Coriat2011} and the majority of BHXBs \citep{Motta2018}. Furthermore, the in-band radio spectral index extracted from the MeerKAT data align well with distribution of (steeper) $\alpha$ observed for other `radio-quiet' sources in the literature \citep{EspinasseFender}.

The ThunderKAT X-ray binary monitoring programme will monitor all bright, active, southern hemisphere XRBs in the radio band for five years. H1743$-$322 is well-known for going into outburst every 6-24 months, and therefore the H1743$-$322 field will likely be revisited. We will then construct the radio and X-ray light curves of this source and all other sources monitored with ThunderKAT to produce one of the largest databases of hard-state XRBs. 

\vspace{-0.5cm}
\section*{Acknowledgements}

This work was supported by the Oxford Centre for Astrophysical Surveys, which is funded through generous support from the Hintze Family Charitable Foundation. We thank the staff at the South African Radio Astronomy Observatory (SARAO) for their rapid scheduling of these observations. The MeerKAT telescope is operated by the South African Radio Astronomy Observatory, which is a facility of the National Research Foundation, an agency of the Department of Science and Innovation. We acknowledge the use of public data from the \textit{Swift} data archive. This research has made use of MAXI data provided by RIKEN, JAXA and the MAXI team. ET and SC acknowledges financial support from the UnivEarthS Labex program of Sorbonne Paris Cite (ANR-10-LABX-0023 and ANR-11-IDEX-0005-02). GRS acknowledges support from an NSERC Discovery Grant (RGPIN-06569-2016). DAHB thanks the National Research Foundation of South Africa for research support.




\vspace{-0.5cm}
\bibliographystyle{mnras}
\bibliography{bib}


\bsp	
\label{lastpage}
\end{document}